\documentclass{mem}
\usepackage{natbib}\usepackage{txfonts}\usepackage{balance}
\usepackage{graphicx}
\usepackage[a4paper,breaklinks,dvipdfm]{hyperref}
\idline{75}{282}
\begin{document}
\def\teff{$T\rm_{eff }$}
\def\kms{$\mathrm {km s}^{-1}$}

\title{
The new technique for the determination of the stellar initial mass function of unresolved stellar populations 
}

   \subtitle{}

\author{
N. \,Podorvanyuk\inst{1}, 
I. \, Chilingarian\inst{2,1}
\and I. \, Katkov\inst{1}
          }


\institute{
Sternberg Astronomical Institute, M.V. Lomonosov Moscow State University, 13 Universitetsky prospect, Moscow, 119992, Russia; email: nicola@sai.msu.ru
\and
Smithsonian Astrophysical Observatory, 60 Garden St MS9 Cambridge MA 02138 USA
}

\authorrunning{Podorvanyuk }

\titlerunning{The new technique for the determination of the IMF}

\abstract{
We present the new technique for the determination of the low-mass slope of the stellar mass function from the pixel-space fitting of integrated light spectra. This technique is the extension of the NBursts full spectral fitting technique. It can be used to constrain the stellar initial mass function (IMF) of compact stellar systems having high relaxation timescales (Hubble time or more). We used Monte-Carlo simulations with mock spectra to test the technique and conclude that: (1) age, metallicity, and IMF can be very precisely determined in the first "unrestricted" variant of the code for high signal-to-noise datasets (S/N=100, R=7000 give the uncertainty of alpha of about 0.1); (2) adding the $M/L$ information significantly improves the precision and reduces the degeneracies, however systematic errors in $M/L$ will translate into offsets in the IMF slope.


}
\maketitle{}

\begin{figure}
\resizebox{\hsize}{!}{\includegraphics[width=15cm,viewport=105mm 10mm 295mm 180mm,clip]{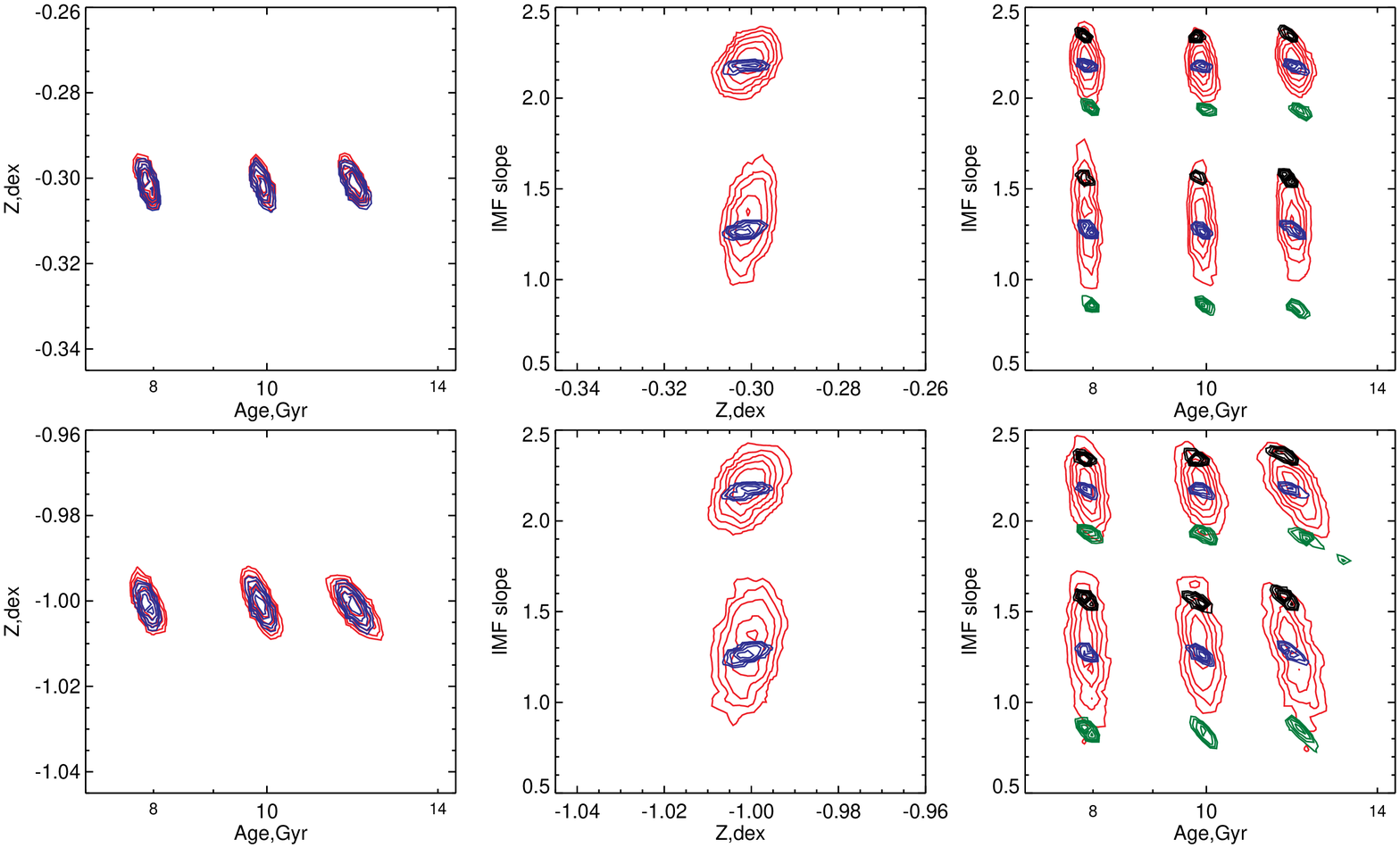}}
\caption{
\footnotesize
Monte Carlo simulations using mock spectra (R=7000) demonstrating the precision of the IMF slope determined by our technique as well as degeneracies between parameters and systematic errors cause by wrong estimates of $(M/L)_{\rm{dyn}}$. Red contours correspond to the results of the fitting by the first version of our technique. The blue contours correspond to the second version with the "external" imposed $M/L$ value. Green and black contours show how the fitting results change when we use the "external" $M/L$ value offset from the input one by $-15$\% and +15\% correspondingly. The degeneracies between metallicity and age (not presented at this figure) are weak (${\Delta}{t}$=0.5 Gyr approximately corresponds ${\Delta}$[Fe/H]=0.05 dex).
}
\label{matrix}
\end{figure}


Globular clusters (GC) and ultracompact dwarf galaxies (UCD) do not have a dark matter according to the modern studies. Hence, their dynamical masses derived from velocity dispersion measurements equal to their stellar masses computed from the stellar population parameters and the slope of the stellar current mass function. If we use the spectra of compact stellar systems having high dynamical relaxation timescales (Hubble time or longer) then the effects of the mass segregation and tidal evaporation of low-mass stars can be neglected and the current stellar mass function should be equivalent to IMF.

Our technique is the extension of the NBursts \citep{chil2007}. NBursts is the approach for determination of parameterized the line of-sight velocity distribution and star formation history. NBursts uses the pre-computed grid of the high-resolution {\sc pegase.hr} \citep{leborgne} simple stellar population (SSP) models.

{\bf In the first version of our new approach} an observed spectrum is fitted against an optimal template represented by a linear combination of SSPs each of them characterised by the age, metallicity and the low-mass IMF slope $\alpha$, determined in the same minimization loop, using 3-dimensional cubic spline interpolation on the grid of high-resolution {\sc pegase.hr} SSP models. We deal with observed spectra of compact stellar systems which can be well reprensented by single component SSP models. The model grid has 25 nodes in age (10~Myr to 20~Gyr), 10 nodes in metallicity ([Fe/H] from -2.5 to 1.0~dex) and 10 nodes in the low-mass IMF slope (0.4 to 3.1 with a step of 0.3).

{\bf The second version of the our technique} is mathematically identical to the original {\sc nbursts} technique but the specific grid of input SSP models is constructed and supplied for every observed spectrum. Here we assume that the object being studied contains no dark matter. In this case, $(M/L)_{\rm{dyn}}$=$(M/L)_{*}$. $(M/L)_{\rm{dyn}}$ were determined from the analysis of their internal structure and observed velocity dispersion profiles and are available in the literature \citep{mieske1, chil2008, chil2010}. $(M/L)_{*}$ can be derived from {\sc pegase.2} stellar population models for every set of (t, [Fe/H], $\alpha$) and for every given t and [Fe/H] in the old stellar population regime this function is monotonic. Hence, if we know $(M/L)_{\rm{dyn}}$ and impose the zero dark matter constraint, for every point on the ($t$, [Fe/H]) plane we can find the value of $\alpha$ such as $(M/L)_{*}$ in that point of the parameter space equals to $(M/L)_{\rm{dyn}}$. Thus, we can compute a grid of SSP models in the age--metallicity space varying the low-mass IMF slope so that the $(M/L)_{*}$ values are constrant all over the grid and equal to the ``external'' $(M/L)_{\rm{dyn}}$. Along with this grid, we will also map the behaviour of $\alpha$ as a function of $t$ and [Fe/H]. Then, if we feed this SSP grid to the standard {\sc nbursts} full spectral fitting procedure and determine the pair of best-fitting values of age and metallicity for some fixed ($M/L$), we will automatically get the $\alpha$ value corresponding to this best-fitting solution which will measure the low-mass IMF slope of the stellar population in the stellar system being studied.

We also applied our technique to observed intermediate and high-resolution GC and UCD spectra obtained with Gemini and VLT. High signal-to-noise spectra having $R=2000$ at $3900<\lambda<6800$~\AA\ \citep{francis} yield to $\sigma(\alpha)=0.2\dots0.4$ and $0.1\dots0.2$ respectively in the first and second verions of our approach.

We conclude that by applying our technique to high-quality optical observations of extragalactic GCs and UCDs we are able to reach better precision of the IMF determination than that made with direct star counts in nearby open clusters and check the IMF universality hypothesis \citep{Kroupa1}.

\bibliographystyle{aa}

\end{document}